\documentclass[prl,superscriptaddress,showpacs,twocolumn,epsf]{revtex4}
\usepackage{epsfig}

\newcommand{\braket}[2]{\langle #1 \,|\, #2 \rangle}
\newcommand{\ket}[1]{| \, #1 \rangle}
\newcommand{\bra}[1]{ \langle #1 \,  |}

\begin{document}
\title{Entangled states that cannot reproduce original classical games \\
in their quantum version}

\author{Junichi Shimamura}
\affiliation{
Graduate School of Engineering Science, Osaka University,
1-3 Machikaneyama, Toyonaka, Osaka 560-8531, JAPAN}
\affiliation{
SORST Research Team for Interacting Carrier Electronics}
\affiliation{CREST Research Team for Photonic Quantum Information}
\author{\c{S}ahin Kaya \"Ozdemir}
\affiliation{
SORST Research Team for Interacting Carrier Electronics}
\affiliation{CREST Research Team for Photonic Quantum Information}
\affiliation{The Graduate University for Advanced Studies
(SOKENDAI), Hayama, Kanagawa 240-0193, Japan}
\author{Fumiaki Morikoshi}
\affiliation{NTT Basic Research Laboratories, NTT Corporation, 3-1 Morinosato
Wakamiya, Atsugi, Kanagawa 243-0198, Japan}
\author{Nobuyuki Imoto}
\affiliation{
SORST Research Team for Interacting Carrier Electronics}
\affiliation{CREST Research Team for Photonic Quantum Information}
\affiliation{The Graduate University for Advanced Studies
(SOKENDAI), Hayama, Kanagawa 240-0193, Japan}
\affiliation{NTT Basic Research Laboratories, NTT Corporation, 3-1 Morinosato
Wakamiya, Atsugi, Kanagawa 243-0198, Japan}
\begin{abstract}
A model of a quantum version of classical games should reproduce the original classical games in order to be able to make a comparative analysis of quantum and classical effects. We analyze a class of symmetric
multipartite entangled states and their effect on the reproducibility
of the classical games. We present the necessary and sufficient
condition for the reproducibility of the original classical games. 
Satisfying this condition means that complete orthogonal
bases can be constructed from a given multipartite
entangled state provided that each party is restricted to two local unitary operators. We prove that most
of the states belonging to the class of symmetric states with
respect to permutations, including the $N$-qubit W
state, do not satisfy this condition.
\end{abstract}
\pacs{03.67.-a, 02.50.Le}
\date{\today}
\pagestyle{plain} \pagenumbering{arabic} \maketitle

{\bf\textit{a. Introduction}:} Entanglement has become
a fundamental ingredient of research in quantum theory
and quantum information processing. Recent studies have brought
this valuable physical resource into the field of game theory,
causing it to emerge as a new field of interest for the quantum
information community
\cite{Eisert1,Meyer,Enk,Du2,Johnson3,Ben,Iqbal,Johnson1}. Quantum versions
of various classical games have been studied and some results
which cannot be seen in the classical versions have been obtained,
i.e., shared entanglement between spatially separated parties can
help them resolve dilemmas in some of the classical games
\cite{Eisert1, samaritan, simamura}.

The effects of different types of entangled states and
their usability in multiplayer multi-strategy games in quantum
settings is still an unexplored area of interest. As an attempt to
fill this gap, we introduce the reproducibility of the classical
version of the game as a criterion to test whether a
given multipartite entangled state can be used in quantum versions
of classical games or not. Any model of a quantum version of given
classical games should include the original classical games as a special
case. In other words, the original classical games should be
reproduced in the model of a quantum version of the classical
games in a certain restricted situation in order to be able to make a
comparative analysis of quantum and classical effects in a game.
This property makes it possible to investigate what is attainable
only in a quantum version.

In classical game theory, a strategic game is defined by
$\Gamma=[N,(S_{i})_{i\in N},(\$_{i})_{i\in N}]$ where $N$ is the
set of players, $S_{i}=\{s_{i}^{1},s_{i}^{2}, \ldots ,
s_{i}^{m}\}$ is the set of pure strategies available to the $i$-th
player with $m$ being the number of strategies, and $\$_{i}$ is
the payoff function for the $i$-th player. When the strategic game
$\Gamma$ is played with pure strategies, the $i$-th player chooses
one of the strategies from the set $S_i$. With all players
applying a pure strategy (each player chooses only one strategy
from the strategy set), the joint strategy of the players is
denoted by $\vec{s}_k=(s_{1}^{l_{1}},s_{2}^{l_{2}}, \ldots,
s_{N}^{l_{N}})$ with $l_{i} = \{ 1,2,3, \ldots ,m \}$ and
$k=\sum_{i=1}^{N}(l_i -1)m^{i-1}$. Then the $i$-th
player's payoff function is represented by $\$_i (\vec{s}_k)$ when
the joint strategy set $\vec{s}_k$ is chosen, i.e., the payoff
functions of all players corresponding to the unique
joint strategy $\vec{s}_k$ can be represented by  $\vec{\$}=(\$_1
(\vec{s}_k),\$_2 (\vec{s}_k), \ldots ,\$_{N} (\vec{s}_k))$ and it
is uniquely determined from the payoff matrix of the game.

So far, most of the studies on quantum versions of
classical games have been based on the model proposed by Eisert
{\it et al.} \cite{Eisert1}. In this model, the strategy set of
the players consists of unitary operators which are applied
locally on a shared entangled state by the players. A measurement
by a referee on the final state after the application of the
operators maps the chosen strategies of the players to their
payoff functions. In this model, the two strategies of
the players in the original classical game is represented by two
unitary operators, $\{\hat{\sigma}_0$,
$i\hat{\sigma}_{y}\}$.


In this paper, we consider the following model of a quantum version of classical games for $N$-player-two-strategy games, which includes Eisert {\it et al.}'s model \cite{Eisert2}: (i) A referee prepares an $N$-qubit
entangled state $\ket{\Psi}$ and distributes it among $N$ players,
one qubit for each player. (In order to see features intrinsic to
quantumness, we focus on a shared entangled state among the
players, and exclude the trivial case where a product state is
distributed.) (ii) Each player independently and locally applies a
unitary operator chosen from the SU(2) set on his
qubit, i.e, the $i$-th player applies $\hat{u}_i$. (We
restrict ourselves to the entire set of SU(2) because the global phase is irrelevant). Hence, the combined strategies of all
the players is represented by the tensor product of all players'
unitary operators as $\hat{x}=\hat{u}_1 \otimes \hat{u}_2 \otimes
\cdots \otimes \hat{u}_N$, which generates the output state
$\hat{x}\ket{\Psi}$ to be submitted to the referee. (iii) Upon
receiving this final state, the referee makes a projective
measurement $\{ \Pi_j \}_{j=1}^{2^{N}}$ which outputs $j$ with probability ${\rm Tr}[\Pi_{j}
\hat{x}\ket{\Psi}\bra{\Psi}\hat{x}^{\dagger}]$, and assigns
payoffs chosen from the payoff matrix, depending on the
measurement outcome $j$.

In this model, we require that a classical game be reproduced when
each player's strategy set is restricted to two unitary operators,
$U_i=\{ \hat{u}_{i}^{1},\hat{u}_{i}^{2} \}$, corresponding to the
two pure strategies in the classical game. When the classical game is
played in this model, the combined strategy of the players is
represented by $\hat{x}_k=\hat{u}_{1}^{l_1} \otimes
\hat{u}_{2}^{l_2} \otimes \cdots \otimes \hat{u}_{N}^{l_N}$ with
$l_i=\{1,2\}$ and $k=\sum_{i=1}^{N}(l_i -1)2^{i-1}$. Thus the
output state becomes $\ket{\Phi_k}=\hat{x}_k\ket{\Psi}$ with $k =
\{1,2, \ldots, 2^N \}$. In order to assign payoffs uniquely, the
referee needs to discriminate all the possible output states
$\ket{\Phi_k}$ deterministically, i.e., the projector $\{ \Pi_j
\}_{j=1}^{2^{N}}$ has to satisfy ${\rm Tr}[\Pi_{j}
\ket{\Phi_k}\bra{\Phi_k}]=\delta_{jk}$. This can be satisfied if
and only if
$\braket{\Phi_\alpha}{\Phi_\beta}=\delta_{\alpha\beta}$ for
$\forall \alpha, \beta$. Therefore, we can say that an entangled
state $\ket{\Psi}$ can be used in this model of quantum version of
classical games if and only if there exist two unitary operators
$\{\hat{u}_{i}^1, \hat{u}_{i}^2\}$ for each player such that the
distinguishability condition,
$\braket{\Phi_\alpha}{\Phi_\beta}=\delta_{\alpha\beta}$ for
$\forall \alpha, \beta$ is satisfied. Otherwise, the players' strategies are mapped to a probability distribution over
the entries of the payoff matrix, which implies that
the original classical game with pure strategies cannot be
reproduced in the quantum game settings.

Any two-qubit pure state satisfies the above
distinguishability condition: The Schmidt decomposition of a
two-qubit state is $\alpha \ket{00} + \beta \ket{11}$ with
$\alpha, \beta$ being real. This state satisfies the
distinguishability condition if the two unitary operators for two
players are chosen as $\{\hat{\sigma}_0,\hat{\sigma}_{x} \}$ and
$\{ \hat{\sigma}_0, i\hat{\sigma}_{y} \}$, respectively. Although
Eisert {\it et al.}'s model was originally proposed for
two-player-two-strategy games, it has recently been extended to
multi-player-two-strategy games by introducing the GHZ-like state
$(\ket{00 \ldots 0} + i \ket{11 \ldots 1})/\sqrt{2}$, which
satisfies the above condition with the two unitary
operators chosen as $\{ \hat{\sigma}_0, i\hat{\sigma}_y \}$ for
each player \cite{Ben}. Of course, any state related
to this state by local unitary transformations also
satisfies the condition. This naturally leads to the
question of whether or not other classes of multipartite entangled states
can satisfy the distinguishability condition. In other
words, we seek for classes of entangled states that
can or cannot be used in the quantum versions of classical games
in this model.

{\bf\textit{b. W state cannot be used}:} The results obtained with
the GHZ-like state encourage us to investigate the W state
\cite{Dur} for multi-player games. We, therefore, consider an $N$-party form of the W state, defined as
$|W_{N}\rangle=|N-1,1\rangle/ {\sqrt{N}}\, (N \ge 3)$, where $|N-1,1\rangle$
is a symmetric state with $N-1$ zeros and $1$ one, e.g. $|2,1\rangle=|001\rangle+|010\rangle+|100\rangle$. 
The reduced density operators of any two-party in the state $\ket{W_N}$ are the
same, and thus entanglement between any two-party is also the same and the maximum possible amount \cite{Koashi}. On the contrary, in the GHZ-like state there is no bipartite entanglement. Consequently, one might expect that using
the state $\ket{W_N}$ can introduce new features into the game.
One possibility might be the case where two of the players decide
to conspire to make their payoffs as high as possible, exploiting
bipartite entanglement shared between them. However, before
analyzing these possibilities, one should be sure that when the
state $\ket{W_N}$ is used in the quantum scheme, the resultant
quantum game should include the original classical game as its subset,
that is, the results of the classical game should be reproducible
in the quantum scheme. In the following, we consider the
distinguishability condition discussed above, for the state
$\ket{W_N}$. In other words, we investigate whether there exist
two unitary operators $\{\hat{u}_i,\hat{v}_i\}$ for each player
such that full orthogonal bases are constructed from the state
$\ket{W_N}$ by applying all possible combinations of the two
unitary operators.

Let us denote the $i$-th player's strategy set
$U_{i}=\{\hat{u}_{i},\hat{v}_{i}\}$ $(i=1,\ldots,N)$ in an
$N$-player game. Since the set of the joint strategies includes
$2^{N}$ elements, these strategies when acted upon the shared
state $\ket{W_N}$ will generate the set of output states
$\{|\Phi_{1}\rangle,|\Phi_{2}\rangle,\ldots
,|\Phi_{2^{N}}\rangle\}$ such that
\begin{eqnarray}\label{N01}
&&|\Phi_{1}\rangle=\left( \hat{u}_{1}\otimes \hat{u}_{2}\otimes \cdots \otimes \hat{u}_{N} \right) \ket{W_N},
\nonumber \\
&&|\Phi_{2}\rangle=\left( \hat{v}_{1}\otimes \hat{u}_{2}\otimes \cdots \otimes \hat{u}_{N} \right) \ket{W_N},
\nonumber \\
&&~~\vdots\nonumber \\
&&|\Phi_{2^{N}}\rangle=\left( \hat{v}_{1}\otimes
\hat{v}_{2}\otimes \cdots \otimes \hat{v}_{N} \right) \ket{W_N}.
\end{eqnarray}
In the following, we prove that there exists no operator set $U_i$
such that
\begin{equation}\label{N02}
\langle
\Phi_{\alpha}|\Phi_{\beta}\rangle=\delta_{\alpha\beta},~~\forall\alpha,\beta\in \{1,\ldots ,2^{N}\}.
\end{equation}

First, let us consider two states in which the
strategies of only one player are different, like $\ket{\Phi_{1}}$
and $\ket{\Phi_{2}}$. The inner product of these states becomes
\begin{equation} \label{N03}
\braket{\Phi_{1}}{\Phi_{2}}=\frac{1}{N}[(N-1)\bra{0}\hat{u}^{\dag}_{1}\hat{v}_{1}\ket{0}
+ \bra{1}\hat{u}^{\dag}_{1}\hat{v}_{1}\ket{1}].
\end{equation}
Since $\hat{u}_{1}$ and $\hat{v}_{1}$ are SU(2) operators, so is $\hat{u}^{\dagger}_{1}\hat{v}_{1}$, which can be 
written as
\begin{equation}\label{N04}
\hat{u}^{\dagger}_{1}\hat{v}_{1}=\left(\begin{array}{cc}
x      & y \\
y^{*} & -x^{*} \\
\end{array} \right)
\end{equation}
with $x$ and $y$ being complex numbers, and $|x|^{2} + |y|^{2} =
1$. Then, Eq. (\ref{N03}) together with the condition
in Eq. (\ref{N02}) implies
\begin{equation}\label{N05}
\hat{u}^{\dagger}_{1}\hat{v}_{1}= \left(
\begin{array}{cc}
0 & e^{i \phi_{1}} \\
 e^{-i \phi_{1}} & 0 \\
\end{array}
\right)
\end{equation}
where $\phi_{1}$ is a real number. Indeed, this holds for all $\hat{u}^{\dagger}_{i}\hat{v}_{i}$ $(i=1,\ldots,N)$.

Next, let us consider the inner product of the following two states,
\begin{eqnarray}\label{N06}
|\Phi_{1}\rangle&=& \left( \hat{u}_{1}\otimes \hat{u}_{2}\otimes \hat{u}_{3}\otimes \cdots \otimes \hat{u}_{N} \right) |W_{N}\rangle,
\nonumber \\
|\Phi_{3}\rangle&=& \left(\hat{v}_{1}\otimes \hat{v}_{2}\otimes
\hat{u}_{3} \otimes \cdots \otimes \hat{u}_{N} \right) |W_{N}\rangle,
\end{eqnarray}
which are different in the strategies of only two
players. The inner product of the states becomes
\begin{eqnarray}\label{N07}
\braket{\Phi_{1}}{\Phi_{3}}&=& \frac{1}{N}\left( \bra{01}+\bra{10}
\right) (\hat{u}^{\dagger}_{1}\hat{v}_{1})
\otimes(\hat{u}^{\dagger}_{2}\hat{v}_{2})
\left( \ket{01}+\ket{10} \right) \nonumber \\
&=& \frac{1}{N}\left( e^{i(\phi_{1}-\phi_{2})} +
e^{-i(\phi_{1}-\phi_{2})} \right) \nonumber\\
&=&  \frac{2}{N} \cos(\phi_{1}-\phi_{2}).
\end{eqnarray}
From the condition in Eq. (\ref{N02}), we obtain
$\phi_{1}-\phi_{2}=\pi/2 + n \pi$ with $n$ being an integer.
Similarly, we have $\phi_{i}-\phi_{j}=\pi/2 + n \pi$ for all
combinations of $i$ and $j$.

Now let us look at the following three cases,
\begin{eqnarray}\label{N08}
\phi_{i}-\phi_{j} &=& \frac{\pi}{2} + n\pi, \nonumber \\
\phi_{j}-\phi_{k} &=& \frac{\pi}{2} + n'\pi, \nonumber \\
\phi_{k}-\phi_{i} &=& \frac{\pi}{2} + n''\pi,
\label{E3:eqn}
\end{eqnarray} where $i\neq j\neq k$, and $n$, $n'$, and $n''$ are integers.
The sum of the above three equations gives $\pi/2=-(n + n' + n'' +
1)\pi$, which is a contradiction. Therefore, there exists no
operator set $U_i$ satisfying the condition of Eq. (\ref{N02}).
This means that the state $\ket{W_N}$ cannot be used \cite{Guo}.
This conclusion is still valid even if a relative phase is added to
each term, because the state $\ket{W_{N}}$ with any relative phase
can be converted into the state $\ket{W_{N}}$ by local unitary
transformations.

{\bf\textit{c. Generalization to Dicke states}:} Next, let us
consider a more general class of symmetric states
$\ket{N-m,m}/\sqrt{{}_{N}C_m}$ with $(N-m)$ zeros and $m$ ones,
which are also known as Dicke states \cite{Mandel}, where
${}_{N}C_m$ denotes the binomial coefficient. The state
$\ket{W_N}$ and the Bell state $(\ket{01} + \ket{10})/\sqrt{2}$
are members of Dicke states, $\ket{N-1,1}/\sqrt{N}$ and
$\ket{1,1}/\sqrt{2}$, respectively. In the following, we will see
that almost all Dicke states cannot be used for the model of a quantum version of $N$-player games as well as the state
$\ket{W_N}$. Note that the GHZ state, which can be used in the model, are not
Dicke states, although they can be represented as superpositions
of two Dicke states $\ket{N,0}$ and $\ket{0,N}$.

Let us consider the inner product of the two output states, like
$\ket{\Phi_1}$ and $\ket{\Phi_2}$, which differ only in
the strategy of one player. The inner product of two such states
$\braket{\Phi_{1}}{\Phi_{2}}$ becomes
\begin{equation}\label{N10}
\frac{N-m}{N}\bra{0}\hat{u}^{\dag}_{i}\hat{v}_{i}\ket{0}
+ \frac{m}{N}\bra{1}\hat{u}^{\dag}_{i}\hat{v}_{i}\ket{1}.
\end{equation}
When the numbers of zeros and ones are different, i.e., $m \neq
N/2$, the coefficients of
$\bra{0}\hat{u}^{\dag}_{i}\hat{v}_{i}\ket{0}$ and
$\bra{1}\hat{u}^{\dag}_{i}\hat{v}_{i}\ket{1}$ are different as in
Eq. (\ref{N03}). Thus, as in the case of the state $\ket{W_N}$,
the diagonal elements of $\hat{u}^{\dag}_i\hat{v}_i$ are zeros as
in Eq. (\ref{N05}). Furthermore, comparing the two output states
in which the strategies of only two players are different, we end
up with the same contradiction as in Eq. (\ref{N08}). Therefore,
we conclude that the class of Dicke states with unequal numbers of
zeros and ones ($\ket{W_N}$ is a member of this class of states)
cannot be used in quantum versions of classical games.

On the other hand, for the Dicke states which contain the same
numbers of zeros and ones, i.e., $\ket{N/2, N/2}$ with an even
number $N$ (the normalization factor is omitted hereafter), we
find that all but $\ket{1,1}$ and $\ket{2,2}$ cannot be used. In
the case of $\ket{2,2}$, an example of the four players' operators
available for the game is
$\hat{u}_{1}=\hat{u}_{2}=\hat{u}_{3}=\hat{u}_{4}=\hat{\sigma}_{0}$,
$\hat{v}_{1}=\hat{v}_{2}=\hat{v}_{3}=(\sqrt{2}\hat{\sigma}_{z}+\hat{\sigma}_{x})/\sqrt{3}$,
and $\hat{v}_{4}=\hat{\sigma}_{y}$.

Next, we prove that the states $\ket{N/2, N/2}$ $(N \ge 6)$ cannot
be used in quantum versions of classical games by showing that
such states cannot satisfy the condition in
Eq. (\ref{N02}). Since $m=N/2$, the inner product in
Eq. (\ref{N10}) gives
\begin{equation}\label{N11}
\hat{u}^{\dag}_{i}\hat{v}_{i}=
\left(\begin{array}{cc}
\cos{\theta_{i}}                   & e^{i \phi_{i}} \sin{\theta_{i}} \\
 e^{-i \phi_{i}} \sin{\theta_{i}}  & - \cos{\theta_{i}}  \\
\end{array} \right)
\end{equation}
where $\theta_i$ and $\phi_i$ are real numbers. Imposing
the condition in Eq. (\ref{N02}) on two output
states which differ only in the strategies of two
players, and using Eq. (\ref{N11}), we obtain
\begin{equation}\label{N12}
\cos{\theta_{i}}\cos{\theta_{j}} = \frac{N}{2} \cos{(\phi_{i} -
\phi_{j})}\sin{\theta_{i}}\sin{\theta_{j}}.
\end{equation}
Similarly, we consider the inner product of two output states in
which the strategies of only four players are different. Together
with Eqs. (\ref{N11}) and (\ref{N12}), the necessary condition of
Eq. (\ref{N02}) gives
\begin{eqnarray}\label{N13}
&&\frac{24}{N(N-2)}\cos{\theta_{i}}\cos{\theta_{j}}\cos{\theta_{k}}\cos{\theta_{l}}\nonumber\\
&&=[\cos\beta_{1}+\cos\beta_{2}+\cos\beta_{3}]\sin{\theta_{i}}\sin{\theta_{j}}\sin{\theta_{k}}\sin{\theta_{l}},\nonumber \\
&&
\end{eqnarray}
where $\beta_{1}=\phi_{i}+\phi_{j}-\phi_{k}-\phi_{l}$,
$\beta_{2}=\phi_{i}-\phi_{j}+\phi_{k}-\phi_{l}$ and
$\beta_{3}=\phi_{i}-\phi_{j}-\phi_{k}+\phi_{l}$. In the following,
we prove that these states cannot be used by showing a
contradiction between Eqs. (\ref{N11}) - (\ref{N13}):

(i) $\theta_{i} \ne n\pi$ for $\forall i$: Assume there exists
$i=i_{0}$ such that $\theta_{i}= n \pi$. Then Eq. (\ref{N12})
implies $\theta_{j}=\pi/2+n'\pi$ for $\forall j\neq i_0$. For any
$j,k \ne i_0$, substituting $\theta_{j,k}=\pi/2+n'\pi$ into Eq.
(\ref{N12}) yields $\phi_{j}-\phi_{k}=\pi/2+m\pi$. These equations
imply a contradiction as in Eq. (\ref{N08}). Therefore $\theta_i$
cannot be $n\pi$ for $\forall i$.

(ii) $\theta_{i} \ne \pi/2 + n \pi$ for $\forall i$: Take an index $i=i_{0}$ such that $\theta_{i}= \pi/2 + n \pi$. Then Eq.
(\ref{N12}) implies two solutions $\theta_{j}=n\pi$, which
is already ruled out in ({\rm i}), and
$\phi_{i_{0}}-\phi_{j}=\pi/2+m\pi$, for $\forall j \neq i_{0}$.
Then for $\forall j,k \ne i_{0}$, we find
$\phi_{j}-\phi_{k}=\pi+m'\pi$. Substituting this in Eq.
(\ref{N12}) gives $\pm (N/2) \sin{\theta_{j}} \sin{\theta_{k}} =
\cos{\theta_{j}}\cos{\theta_{k}}$. Using these equalities in Eq.
(\ref{N13}) together with $\theta_{i}\neq n\pi$, we find
${6N}/(N-2)=\pm[\cos\beta_{1}+\cos\beta_{2}+ \cos\beta_{3}]$. This
implies a contradiction because while the right hand side lies in
the range $[-3,3]$, the left hand side is greater then 6 as $N \ge
6$. Therefore $\theta_{i}$ cannot be $ \ne \pi/2 + n \pi$ for
$\forall i$.

Applying ({\rm i}) and ({\rm ii}) to Eqs.(\ref{N12}) and (\ref{N13}) completes the proof: Let us take four
different indices $i,j,k$ and $l$. Multiplying two equations
obtained from Eq. (\ref{N12}) for the index pairs $(i,j)$ and
$(k,l)$ yields
$\cos{\theta_{i}}\cos{\theta_{j}}\cos{\theta_{k}}\cos{\theta_{l}}=(N^{2}/4)[\cos{(\phi_{i}-\phi_{j})}\cos{(\phi_
{k}-\phi_{l})}]\sin{\theta_{i}}\sin{\theta_{j}}\sin{\theta_{k}}\sin{\theta_{l}}$.
Comparison of these equations for different pairs of the indices,
such as $(i,k)$ and $(j,l)$, together with the condition
$\theta_{i,j,k,l} \ne n\pi$ reveals
$\cos{(\phi_{i}-\phi_{j})}\cos{(\phi_{k}-\phi_{l})}=\cos{(\phi_{i}-\phi_{k})}\cos{(\phi_{j}-\phi_{l})}=\cos{(\phi_
{i}-\phi_{l})}\cos{(\phi_{j}-\phi_{k})}$.
Then the trigonometric identity $2\cos a\cos
b=\cos(a+b)+\cos(a-b)$ leads to
$\cos\beta_{1}=\cos\beta_{2}=\cos\beta_{3}$. From the first
equation we have $\beta_{1}=\mp \beta_{2}+2m\pi$. Now let us
consider the first case $\beta_{1}=\beta_{2}+2m\pi$, which is
simplified to $\phi_{j}-\phi_{k}=m\pi$. From Eqs. (\ref{N12}) and
(\ref{N13}), we have
$(2N/(N-2))\cos{(\phi_{i}-\phi_{l})}\cos{(\phi_{j}-\phi_{k})}=\cos\beta_{1}$.
Substituting $\phi_{j}-\phi_{k}=m\pi$ in both sides of the
equation, we obtain $\pm(2N/(N-2))\cos{(\phi_{i} -
\phi_{l})}=\cos{(\phi_{i}-\phi_{l})}$. Since $N \ge 6$, this
equation is satisfied only if $\phi_{i}-\phi_{l}=\pi/2+m'\pi$.
Substituting this in Eq. (\ref{N12}) yields $\theta_i = \pi /2 +
n'\pi$ or $\theta_l = \pi /2 + n'\pi$, which contradicts the fact
proven in ({\rm ii}) that $\theta_i \ne \pi/2 + n\pi$ for $\forall
i$. It is easily seen that the same argument holds for the case
$\beta_{1}=-\beta_{2}+2m\pi$. This completes the proof that the
state $\ket{N/2,N/2}$ for $(N \ge 6)$ cannot be used in quantum
versions of classical games in this specific model.

{\bf\textit{d. Conclusion}:} We have investigated whether any
multipartite entangled states can be used in our model
of a quantum version of classical games, which includes Eisert {\it et al.}'s model. We have 
presented the necessary and sufficient condition for the
reproducibility of classical games in this model. This
study reveals that none of the Dicke
states of the form $\ket{N-m,m}$, except $\ket{1,1}$ and
$\ket{2,2}$, can satisfy this condition.
Our argument on the reproducibility of classical games can be
rephrased as the possibility of constructing complete orthogonal
bases from an initial state by applying all possible combinations
of two local unitary operations on each qubit. It is naturally
possible to construct the bases from $N$-qubit product states. In
entangled states, however, this is not always true, i.e., while
this task is possible starting with an $N$-qubit GHZ state $(\ket{00
\ldots 0} + \ket{11 \ldots 1})/\sqrt{2}$, it is impossible when
the $N$-qubit state is the state $\ket{W_N}$.
This is another difference in the properties
of the GHZ and the W state.
We think our analysis might be useful in understanding the nature of multipartite entanglement in quantum information processing. The study on
 how this property of entanglement is related to other quantum information processing tasks remains as a subject to be discussed further.

\begin{acknowledgments}
The authors thank M. Koashi for helpful discussions and critical reading 
of the manuscript, and T. Yamamoto and J. Soderholm for their warm support during this research.
This work is partially supported by a Grant-in-Aid for Scientific 
Research (C) (15300079) by the Japan Society for the Promotion of 
Science (JSPS)
and by the Telecommunications Advancement Organization of Japan (TAO) .
\end{acknowledgments}

\end{document}